\documentclass[12pt]{article}
\usepackage{max-dem}
\begin{document}
\title{Maxwell Demon from a\\
Quantum Bayesian Networks
\\Perspective}

\author{Robert R. Tucci\\
        P.O. Box 226\\
        Bedford,  MA   01730\\
        tucci@ar-tiste.com}

\date{\today}
\maketitle
\vskip2cm
\section*{Abstract}
We propose a new inequality
that we call the
conditional ageing
inequality (CAIN). The CAIN
is a slight generalization
to non-equilibrium situations
of the Second Law of thermodynamics.
The goal of this paper is
to study the consequences of
the CAIN.
We use the CAIN to
discuss
Maxwell demon processes
(i.e., thermodynamic
processes with feedback.)
 In particular,
we apply the CAIN to four
cases of the Szilard engine:
for a classical or a quantum
system with either one or two
correlated particles.
Besides proposing this new
inequality that we call the CAIN,
another novel feature of
this paper is that we
use quantum Bayesian networks
for our analysis
of Maxwell demon processes.

\newpage

\section{Introduction}

In Ref.\cite{Max}, Maxwell proposed
his famous
gedanken experiment
wherein a demon controls the flow
of gas particles from one chamber to another
and decides which particles
to let through based on their temperature.
He gave this thought experiment
as an example of a thermodynamic process
in which the Second Law of
thermodynamics
appears to be violated.
He dismissed
the paradox by saying
that the Second Law
is true only on average.
In Ref.\cite{Szi}, Szilard proposed
an engine which is a simplified version
of a Maxwell's demon. Szilard
argued that for his engine,
the Second Law
is not violated at all, as long as the work
performed by the demon to
make his measurements is taken into
account.
In Refs.\cite{Lan} and \cite{Ben},
 Landauer and later Bennett pointed out
that measurements can
be performed without spending any energy,
but that in order for an engine to perform
a cyclic process, it needs to
store the information of the measurement
on a tape and then erase and re-initialize that
tape once per cycle. These tape operations
will always consume an amount of
energy
larger or equal to
the work the demon can extract
from changes in gas volumes.

Maxwell's demon thought
experiment might have once
been considered paradoxical,
but after the work of
Szilard, Landauer and Bennett,
most scientists consider
the paradox pretty much solved.
Nevertheless,
some people, myself included,
still strive to
make the mathematics involved
in the treatment of Maxwell's demon
a bit more streamlined. That is one of
the goals of this paper,
to look at Maxwell's demon
from a different
point of view,
hoping
that this might yield new
insights to an
already understood problem.

This paper originated as an
attempt to understand  a series of
papers (Refs.\cite{Sag-U-07}
to \cite{Sag-U-12b}) by Sagawa, Ueda
and coworkers (S-U) in which they
claim that
the standard Second Law of thermodynamics
does not apply
to non-equilibrium processes
with feedback
(i.e., Maxwell demon type
processes).
They give a generalization
of the Second Law that
they claim does
apply to such processes.
Although I agree in
 spirit with much of what
S-U are trying
to do, and I profited
immensely from reading their papers,
I disagree with
some of the details of their theory.
I discuss my disagreements
with the S-U theory
in a separate paper,
Ref.\cite{Tuc-disagrees}.
The goal of this paper is
to report on my own theory
for generalizing the Second Law
so that is applies to processes with
feedback.
My theory agrees in spirit with
the S-U theory,
but differs from it in some important
details.

Let me explain the
rationale behind my theory.

Suppose we
want to consider a
system
in thermal contact
but not necessarily in equilibrium with
a bath at temperature $T$.
Let
$\rvX$ denote all
non-thermal variables (fast changing, not in thermal equilibrium)
and let
$\rvTh$ denote
all thermal variables (slow changing, in thermal equilibrium)
describing both the system and bath.
Let $\tau$ denote time.
For any operator $\Omega_\tau$,
define
$\Omega_\tau\trange{\tau_1}{\tau_2}=
\Omega_{\tau_2}-\Omega_{\tau_1}
$. My slight
generalization of the Second Law is

\beq
\fbox{$\displaystyle
\;\;S_\tau(\rvTh_\tau|\rvX_\tau)\trange{0}{\tau}\geq 0
\;\;$}
\label{eq-cain}
\;,
\eeq
where $S(\rva|\rvb)$ is the conditional entropy
(i.e., conditional spread)
of $\rva$ given $\rvb$.
I call Eq.(\ref{eq-cain}) the conditional ageing
inequality (CAIN).
The standard Second Law corresponds
to the special case when
there are no $\rvX_\tau$ variables,
in which case
Eq.(\ref{eq-cain}) reduces to

\beq
S_\tau(\rvTh_\tau)\trange{0}{\tau}\geq 0
\label{eq-sec-law}
\;.
\eeq
The standard Second Law could be described
as unconditional ageing, or simply as ageing.

Now, what is the justification
for the CAIN? The justification
for the Second Law
Eq.(\ref{eq-sec-law})
is that
the superoperator
that evolves the overall probability
distribution in the classical
case (or
the overall density matrix
in the quantum case),
from time 0 to $\tau$,
increases entropy
because
it can be shown to
be doubly stochastic
in the classical case (or
unital\footnote{A
superoperator is unital if
it maps the identity matrix
to itself.} in the quantum case).
The justification for
the CAIN is the same,
except that the evolution
superoperator
is doubly stochastic (or unital)
only if the non-thermal
variables are held fixed during
the evolution.
The CAIN is not true
for all evolution superoperators.
Our hope is
that it applies to
systems of interest
that commonly occur in nature.

The goal of this paper is
to study the consequences of
the CAIN. In particular,
we apply the CAIN to four
cases of the Szilard engine:
for a classical or a quantum
system with either one or two
correlated particles.

Besides proposing this new
inequality that we call the CAIN,
another novel feature of
this paper is that we
use quantum Bayesian networks
for our analysis
of Maxwell demon type processes.

This paper is written assuming that
the reader has first read
Refs.\cite{Tuc-mixology}
and \cite{Tuc-ineq}.
Ref.\cite{Tuc-mixology}
is an introduction
to quantum Bayesian networks for mixed states.
Ref.\cite{Tuc-ineq}
discusses well-known
inequalities of classical
and quantum SIT (Shannon Information Theory)
from
a Bayesian networks perspective.

In this paper, we will use
the abbreviation $\frac{f(a)}{\sum_a(num)}=
\frac{f(a)}{\sum_a f(a)}$.
We will also
use the abbreviations
$\Gamma_{a\;:\;b} = (\Gamma_a,\Gamma_{a+1}, \Gamma_{a+2},
 \ldots, \Gamma_b)$
 and
 $\Gamma_{<b}=\Gamma_{0\;:\;b}$, for
 any vector $\Gamma_a$
and any integers $a, b$
such that $0\leq a\leq b$.

\section{Review of Some Properties of \\
Thermal States}

In this section, we will
review some well known properties of
thermal states
that we shall use later on in the paper
to study some consequences of
the CAIN.
Most of the contents of this section
can be found in reviews about entropy
such as Ref.\cite{Wehrl} by Wehrl and
textbooks on Statistical Mechanics such as
Ref.\cite{Fey} by Feynman.

Suppose $\rvj$ is a
classical random variable
that can take on values $j\in S_\rvj$
and has a
probability distribution $P_\rvj(j)$.
We will denote the average
of any function $f:S_\rvj\rarrow \RR$ by

\beq
\av{f(j)}_j =
\av{f(j)}_{P_\rvj}=
\sum_j P(j) f(j)
\;.
\eeq

When speaking about quantum physics,
if $\rho$ is a density operator
acting on a Hilbert space $\hil$,
and $\Omega$ is a Hermitian operator
also acting on $\hil$,
we will denote the average of
$\Omega$ by

\beq
\av{\Omega}_\rho=\tr(\rho\Omega)
\;.
\eeq
For example,
in this notation the von Neumann entropy
of $\rho$ is

\beq
S(\rho) = -\av{\ln\rho}_\rho
\;.
\label{eq-von-neu-ent}
\eeq

Consider a system with density matrix $\rho$
and Hamiltonian
$\ham$. Suppose  the eigenvalue decomposition
 of $\ham$
is $\ham = \sum_j P(j)
 \ket{E_j}\bra{E_j}$.
The internal energy
of the system is defined as

\beq
U = E = \av{\ham}_\rho = \av{E_j}_j
\;.
\eeq

\subsection{Simple
Properties of Thermal States}

Thermal states (a.k.a canonical ensemble
or Gibbs states)
are states with a definite
temperature $T$. Their
form is given below.

In this paper, we
will use what are called natural
Planck units. As
in Eq.(\ref{eq-von-neu-ent}),
our entropies will be
defined in
terms of natural
logs (instead of base 2 logs)
and without the $k_B$.
($k_B$ is Boltzmann's constant.)
Temperatures will be given
in energy units and entropies in nats.
If $T$
is the temperature in energy
units and $T^{Kel}$
is the temperature in degrees Kelvin,
then $T = k_B T^{Kel}$.
We will also use $\beta = \frac{1}{T}$.

Consider a system with Hamiltonian
$\ham = \sum_j P(j)
 \ket{E_j}\bra{E_j}$.
which has reached thermal equilibrium
at a temperature $T$.
The partition function
of the system is defined by

\beq
Z^\beta(\ham) = \tr (e^{-\beta \ham})
=\sum_j e^{-\beta E_j}
\;.
\eeq
Its density matrix is

\beq
\rho^\beta(\ham) =
\frac{R}{Z^\beta(\ham)}
\;, \mbox{ where }R = e^{-\beta \ham}
\;.
\eeq
Its entropy is

\beq
S^\beta(\ham) = S(\rho^\beta(\ham))
\;.
\eeq
Its free energy is

\beq
F^\beta(\ham) = -T\ln Z^\beta(\ham)
\;.
\eeq
Its pressure $P$ (not to be confused with
probability $P(j)$) is

\beq
P= -\av{\left(\pder{E_j}{V}\right)_T}_j
\;.
\eeq
Later we will
show that
this expression for pressure
gives the expected $dE=-PdV$
(Thus, internal energy
of system decreases if system does
work by increasing its volume by $dV$).

\begin{claim}
Let
$S=S^\beta(\ham)$,
$E=\av{\ham}_{\rho^\beta(\ham)}$,
and
$F = F^\beta(\ham)$. Then
\begin{subequations}
\beq
E = TS + F
\;.
\label{eq-u-ts-plus-f}
\eeq
(Thus, internal energy is sum of bound part ($T$ times
entropy)
and free part (free energy)). Furthermore

\beq
\left(\pder{F}{V}\right)_T = -P,
\;\;
\left(\pder{F}{T}\right)_V = -S
\;.
\label{eq-pder-f}
\eeq
(Thus, free energy decreases if
volume or temperature increase). Furthermore

\beq
dF = -SdT -PdV
\;,
\label{eq-df}
\eeq
and

\beq
dE=TdS - PdV
\;.
\label{eq-du}
\eeq
\end{subequations}
\end{claim}
\proof

To prove Eq.(\ref{eq-u-ts-plus-f}), note that

\beq
S
=
\av{\ln\frac{Z}{e^{-\beta E_j}}}_j
=
\ln Z + \beta \av{E_j}_j
=
-\frac{F}{T}
+\frac{E}{T}
\;.
\eeq

To prove Eq.(\ref{eq-pder-f}), note that

\beq
\left(\pder{F}{V}\right)_T
=
-T\left(\pder{\ln Z}{V}\right)_T
=
\frac{-T}{Z}
\sum_j (-\beta e^{-\beta E_j})
\left(\pder{E_j}{V}\right)_T
= -P
\;.
\eeq

\beqa
\left(\pder{F}{T}\right)_V
&=&
-\ln Z - \frac{T}{Z}
\left(\pder{Z}{T}\right)_V
\\
&=&
-\ln Z +
\beta\left(\pder{\ln Z}{ \beta}\right)_V
\\
&=&
-\ln Z -E\beta = -S
\;.
\eeqa

To prove Eq.(\ref{eq-df}), note that

\beq
dF =
\left(\pder{F}{V}\right)_T dV
+
\left(\pder{F}{T}\right)_V dT
\;.
\eeq

To prove Eq.(\ref{eq-du}),
just use
Eqs.(\ref{eq-u-ts-plus-f}) and (\ref{eq-df}).
\qed

\begin{claim}
$S^\beta(\ham)$ and
$\av{\ham}_{\rho^\beta(\ham)}$
are monotonically increasing and
$F^\beta(\ham)$ is monotonically
decreasing
functions of temperature. In fact,

\beq
-\frac{dS(\rho)}{\beta d\beta}=
-\frac{d\av{\ham}_\rho}{d\beta}=
\av{(\ham -\av{\ham}_\rho)^2}_\rho\geq 0
\;,
\eeq
and
\beq
\frac{dF^\beta(\ham)}{d\beta}
= \frac{S(\rho)}{\beta^2}
\;,
\eeq
where we are abbreviating $\rho^\beta(\ham)$
by just $\rho$.
\end{claim}
\proof
Just straightforward Calculus.
\qed

\begin{claim}
Let
$S=S^\beta(\ham)$,
$E=\av{\ham}_{\rho^\beta(\ham)}$,
and
$F = F^\beta(\ham)$. Then

\beq
\begin{array}{|c|c|c|}
\hline
&T\rarrow 0 &T\rarrow \infty\\
\hline\hline
S= & 0& \ln N\\
\hline
E= & E_0 & \frac{1}{N}\sum_j E_j\\
\hline
F=& E_0 & -T\ln N\\
\hline
\end{array}
\;,
\eeq
where $\{E_j\}_{j=0}^{N-1}$
are the eigenvalues of $\ham$
and $E_0$ is the lowest one.
\end{claim}
\proof
Obvious.
\qed

\subsection{Inequalities Relating a
Thermal State With a Neighboring State}

Consider any Hilbert space
$\hil$, any density matrix $\rho$
acting on $\hil$,
any
Hamiltonian $\ham$ acting on $\hil$,
and any temperature $T$.
Define

\beq
S^\beta(\ham,\rho)
= \beta[\av{\ham}_\rho - F^\beta(\ham)]
\;
\eeq
and

\beq
F^\beta(\ham,\rho)
= \av{\ham}_\rho - TS(\rho)
\;.
\eeq
I will refer these functions
as the $S$ and $F$ {\it capping functions},
respectively, because,
as we will prove later, they are
upper bounds to their namesakes.

It's easy to check that
$S^\beta(\ham,\rho^\beta(\ham) )
= S^\beta(\ham)$ and
$F^\beta(\ham, \rho^\beta(\ham))
= F^\beta(\ham)$.

\begin{claim}
\beqa
D(\rho//\rho^\beta(\ham))
&=&
S^\beta(\ham, \rho)-S(\rho)
\label{eq-d-rho-over-rho-beta-s}
\\
&=&
\beta[
F^\beta(\ham,\rho)-F^\beta(\ham)]
\label{eq-d-rho-over-rho-beta-f}
\;.
\eeqa
\end{claim}
\proof
\beqa
D(\rho//\rho^\beta(\ham))&=&
\av{\ln \rho - \ln\rho^\beta(\ham)}_\rho
\\
&=&
\beta[-TS(\rho)+\av{\ham}_\rho -
F^\beta(\ham)]\\
&=&
S^\beta(\ham, \rho)-S(\rho)\\
&=&
\beta[
F^\beta(\ham,\rho)-F^\beta(\ham)]
\;.
\eeqa
\qed

\begin{claim}
\beq
S(\rho)\leq S^\beta(\ham, \rho)
\;.
\label{eq-s-leq-upa}
\eeq
If
$\av{\ham}_\rho\leq
\av{\ham}_{\rho^\beta(\ham)}$,
then also

\beq
S(\rho)\leq S^\beta(\ham)
\;.
\label{eq-s-leq-s-beta}
\eeq
(Eq.(\ref{eq-s-leq-s-beta}) agrees with
our intuition that
$\av{\ham}_\rho$ and $S(\rho)$
both measure the energy spread of $\rho$.)
\end{claim}
\proof
Eq.(\ref{eq-s-leq-upa})
 follows from Eq.(\ref{eq-d-rho-over-rho-beta-s}).

If
$\av{\ham}_\rho\leq
\av{\ham}_{\rho^\beta(\ham)}$,
then
\beqa
S(\rho)&\leq& \beta[\av{\ham}_\rho
- F^\beta(\ham)]
\\
&\leq&
\beta[\av{\ham}_{\rho^\beta(\ham)}
- F^\beta(\ham)]
\\
&=& S^\beta(\ham)
\;.
\eeqa
\qed

\begin{claim}
\beq
F^\beta(\ham)\leq F^\beta(\ham,\rho)
\;.
\label{eq-f-leq-upa}
\eeq
Also

\beq
F^\beta(\ham)\leq \av{\ham}_\rho
\;.
\label{eq-no-free-lunch}
\eeq
Thus, the free energy is always less
than the average energy. (There is no free lunch.)
\end{claim}
\proof
Eq.(\ref{eq-f-leq-upa}) follows from
Eq.(\ref{eq-d-rho-over-rho-beta-f}).

Eq.(\ref{eq-no-free-lunch})
follows from Eq.(\ref{eq-s-leq-upa})
 and the definition of $S^\beta(\ham, \rho)$.
\qed

Suppose $\ham_1$ and $\ham_2$
are two Hamiltonians acting on the same
Hilbert space.
If $[\ham_1,\ham_2]=0$, then
clearly $Z^\beta(\ham_1+\ham_2)=
Z^\beta(\ham_1)Z^\beta(\ham_2)$
so $F^\beta(\ham_1+\ham_2)= F^\beta(\ham_1)
+ F^\beta(\ham_2)$. But what if
$\ham_1$ and $\ham_1$ don't commute?
Is the free energy sub-additive or
super-additive
(or neither) in its Hamiltonian?

\begin{claim}(Peierls-Bogoliubov)\footnote{This
inequality is referred to as the
Peierls-Bogoliubov inequality
in the review by Wehrl\cite{Wehrl}.
It's used in Feynman's Statistical Mechanics\cite{Fey}
book to do variational approximations of
the free energy. As shown here,
it follows trivially from the
monotonicity of the relative entropy,
which was found by Uhlmann and others.}

\beq
F^\beta(\ham_2)\leq
F^\beta(\ham_1) +
\av{\ham_2-\ham_1}_{\rho^\beta(\ham_1)}
\;.
\label{eq-pie-bog-ineq}
\eeq
\end{claim}
\proof

\beqa
D(\rho^\beta(\ham_1)//\rho^\beta(\ham_2))
&=&
\av{\ln \rho^\beta(\ham_1)-
\ln \rho^\beta(\ham_2)}_{
\rho^\beta(\ham_1)}
\\
&=&
-S^\beta(\ham_1)
+\beta\av{\ham_2}_{\rho^\beta(\ham_1)}
-\beta F^\beta(\ham_2)
\\
&=&
-S^\beta(\ham_1)
+\beta\av{\ham_1}_{\rho^\beta(\ham_1)}
-\beta F^\beta(\ham_2)
+\beta\av{\ham_2-\ham_1}_{\rho^\beta(\ham_1)}
\\
&=&
\beta [F^\beta(\ham_1)-F^\beta(\ham_2)]
+\beta\av{\ham_2-\ham_1}_{\rho^\beta(\ham_1)}
\;.
\eeqa
\qed

\begin{claim}
\beq
F^\beta(\ham)+\av{\Delta\ham}_{\rho(\ham+\Delta\ham)}
\stackrel{(a)}{\leq}
F(\ham+ \Delta \ham)
\stackrel{(b)}{\leq}
F^\beta(\ham)+\av{\Delta\ham}_{\rho(\ham)}
\;.
\label{eq-free-u-and-l-bound}
\eeq
\end{claim}
\proof

Inequality $(a)$ follows if one
sets $\ham_1=\ham+\Delta\ham$ and $\ham_2=\ham$
in Eq.(\ref{eq-pie-bog-ineq}).

Inequality $(b)$ follows if one
sets $\ham_1=\ham$ and $\ham_2=\ham+\Delta\ham$
in Eq.(\ref{eq-pie-bog-ineq}).
\qed

\begin{claim}
\beq
F^\beta(\ham) +F^\beta(\Delta\ham)
\leq
F^\beta(\ham+\Delta\ham)
\;.
\eeq
\end{claim}
\proof
Just use the no-free lunch inequality
in Eq.(\ref{eq-free-u-and-l-bound}) side $(a)$.
\qed

\section{The Conditional Ageing Inequality
and Some of its Consequences}

In Appendix \ref{app-thermo},
we reminded the reader of
the well know inequality $dW_s\leq -dF_s$,
which says that
at fixed temperature,
 the drop in free energy
is an upper bound to the amount of
work system $s$ can do.
In this section we apply
the conditional ageing inequality
to find: a
lower bound on $-dF_X$ for
a system $\rvX$ in contact
with a heat reservoir $\rvTh$
at temperature $T$.

We will abbreviate
$\rho_{\tau;\rvTh_\tau, \rvX_\tau}$ by
$\rho_\tau$.
The partial traces of
$\rho_{\tau;\rvTh_\tau, \rvX_\tau}$
will be denoted by
$\rho_{\tau;\rvTh_\tau}$
and
$\rho_{\tau;\rvX_\tau}$.
We will also abbreviate
$S_\tau(\cdot)= S_{\rho_{\tau;\rvTh_\tau,
\rvX_\tau}}(\cdot)$
for any argument $(\cdot)$.

Let the joint system of $\rvX$ and $\rvTh$ have
as Hamiltonian
\beq
\ham_{\rvTh_\tau,\rvX_\tau}=
\ham_{\rvX_\tau} +
\ham_{\rvTh_\tau} +
\epsilon_{\rvTh_\tau, \rvX_\tau}
=
\ham_{\rvX_\tau} +
\Delta \ham_{\rvTh_\tau,\rvX_\tau}
\;,
\eeq
where $[\ham_{\rvX_\tau},
\ham_{\rvTh_\tau}]=0$ and
$\epsilon_{\rvTh_\tau, \rvX_\tau}$ is
small.

The conditional ageing inequality (CAIN)
is

\beq
S_\tau(\rvTh_\tau|\rvX_\tau)\trange{0}{\tau}\geq 0
\;.
\eeq
Besides the CAIN,
we will also assume
that the following is true at $\tau=0$:
$\rvTh_0$ and
$\rvX_0$
are independent
and thermal.
The independence is achieved by assuming that
$\epsilon_{\rvTh_0,\rvX_0}=0$.

\begin{claim}
If the CAIN holds,
and
$\rvTh_0$ and
$\rvX_0$
are independent
and thermal,
then
\beq
-F^\beta(\ham_{\rvX_\tau},
 \rho_{\tau;\rvX_\tau})
\trange{0}{\tau}
-TS^\beta(\Delta \ham_{\rvTh_\tau,\rvX_\tau}, \rho_\tau)
\trange{0}{\tau}
\leq
-F^\beta(\ham_{\rvX_\tau})\trange{0}{\tau}
\;
\label{eq-lower-bd-minus-df}
\eeq
where

\beq
-F^\beta(\ham_{\rvX_\tau})
=
TS^\beta(\ham_{\rvX_\tau})
-
\av{\ham_{\rvX_\tau}}_{\rho^\beta(\ham_{\rvX_\tau})}
\;
\eeq
and
\beq
-F^\beta(\ham_{\rvX_\tau},
 \rho_{\tau;\rvX_\tau})
=
TS(\rho_{\tau;\rvX_\tau})
-
\av{\ham_{\rvX_\tau}}_{\rho_{\tau;\rvX_\tau}}
\;.
\label{eq-minus-f-rho-x}
\eeq
\end{claim}
\proof

The CAIN implies
\beq
S_\tau(\rvX_\tau)\trange{0}{\tau}\leq
S_\tau(\rvTh_\tau,\rvX_\tau)\trange{0}{\tau}
\;.
\label{eq-cain-long}
\eeq
But

\beqa
S_\tau(\rvTh_\tau,\rvX_\tau)
&=&
S(\rho_{\tau;\rvTh_\tau,\rvX_\tau})
\\
&\leq&
\beta\left[
\av{\Delta \ham_{\rvTh_\tau,\rvX_\tau} +
\ham_{\rvX_\tau}}_{\rho_\tau}
-
F^\beta(\Delta \ham_{\rvTh_\tau,\rvX_\tau} +
\ham_{\rvX_\tau})
\right]
\\
&\leq&
\beta\left[
\av{\Delta \ham_{\rvTh_\tau,\rvX_\tau} +
\ham_{\rvX_\tau}}_{\rho_\tau}
-
F^\beta(\Delta \ham_{\rvTh_\tau,\rvX_\tau}) -
F^\beta(\ham_{\rvX_\tau})
\right]
\\
&=&
S^\beta(\Delta \ham_{\rvTh_\tau,\rvX_\tau}, \rho_\tau) +
S^\beta(\ham_{\rvX_\tau}, \rho_\tau)
\;.
\label{eq-s-tx-tau}
\eeqa
Also, since $\rvX_0$ and $\rvTh_0$
are independent and thermal,

\beqa
S_0(\rvTh_0,\rvX_0)&=&
S^\beta(\ham_{\rvTh_0}) +
S^\beta(\ham_{\rvX_0})
\\
&=&
S^\beta(\ham_{\rvTh_0}, \rho_0) +
S^\beta(\ham_{\rvX_0},\rho_0 )
\;.
\label{eq-s-tx-zero}
\eeqa
Combining Eqs.(\ref{eq-cain-long}),
(\ref{eq-s-tx-tau})
and (\ref{eq-s-tx-zero}) yields

\beq
S(\rho_{\tau;\rvX_\tau})\trange{0}{\tau}
=
S_\tau(\rvX_\tau)\trange{0}{\tau}
\leq
S^\beta(\Delta \ham_{\rvTh_\tau,\rvX_\tau},\rho_\tau )\trange{0}{\tau}
+
S^\beta(\ham_{\rvX_\tau},\rho_\tau )\trange{0}{\tau}
\;.
\eeq
Now using

\beq
S^\beta(\ham_{\rvX_\tau},\rho_\tau
)\trange{0}{\tau}
=\beta[
\av{\ham_{\rvX_\tau}}_{\rho_\tau}
-F^\beta(\ham_{\rvX_\tau})]\trange{0}{\tau}
\;
\eeq
gives

\beqa
-\beta F^\beta(\ham_{\rvX_\tau},\rho_{\tau;\rvX_\tau} )
\trange{0}{\tau}
&=&
S(\rho_{\tau;\rvX_\tau})\trange{0}{\tau}
-
\beta\av{\ham_{\rvX_\tau}}_{\rho_\tau}\trange{0}{\tau}
\\
&\leq&
S^\beta(\Delta \ham_{\rvTh_\tau,\rvX_\tau}, \rho_\tau)\trange{0}{\tau}
-\beta F^\beta(\ham_{\rvX_\tau})\trange{0}{\tau}
\;.
\eeqa
\qed

\section{Conditional Ageing
in Terms of Time Reversal}

In this section, we will
state the CAIN in terms of
time reversal.
The Second Law of
Thermodynamics
and it's generalization,
the Jarzynski identity\cite{Jar},
are often stated
using time reversal ideas.
This is a natural thing to do
since they both describe
entropy changes and such changes
arises from
irreversible processes.
The CAIN can be viewed
as a slight
generalization of
the Second Law,
so it too should be stateable in
terms of time reversal.

For a good
pedagogical treatment of
time reversal, see,
for example, Ref\cite{Joshi}.

In classical physics,
given a system of $N$ particles
labeled by $\mu=1,2, \ldots, N$,
if
$f(\{
\vec{r}_\mu,
\vec{p}_\mu
\}_{\forall \mu})$
is a function of the
positions $\vec{r_\mu}$
and momenta $\vec{p_\mu}$ of
the particles,
then
the time reversal operator,
which we will represent by
$\trev$, keeps the position vectors
the same, but it reverses the velocities,
and therefore the momenta.
Thus
$[f(\{
\vec{r}_\mu,
\vec{p}_\mu
\}_{\forall \mu})]^\trev=
f(\{
\vec{r}_\mu,
-\vec{p}_\mu
\}_{\forall \mu})$.

In quantum mechanics,
if we express
all operators
and wavefunctions
in
position and spin space,
then
$[f(\{
\vec{r}_\mu,
\vec{p}_\mu
\}_{\forall \mu})]^\trev=
f(\{
\vec{r}_\mu,
-\vec{p}_\mu
\}_{\forall \mu})$
still applies,
where now $f$
is either
an observable or
a wavefunction.
The
position operators
$\vec{r}_\mu$ are
real and the momentum
operators
$\vec{p}_\mu$
are pure imaginary.
Thus,
in the case of spinless particles,
$\trev$ can be taken
to be simply complex conjugation $*$.
If the particles do have spin,
then one must also
rotate the spin space part of $f(\cdot)$
by a matrix which is real,
and therefore commutes
with complex conjugation.
See Ref.\cite{Joshi} for more
details on how to deal with spin.
In this
paper, we will only discuss the
spinless case.

This paper is mainly concerned with
 the time reversal of a
simple Markov
chain.
For example, in later sections of the paper,
we will model the classical Szilard engine
by a
CB net of the form

\beq
\entrymodifiers={++[o][F-]}
\xymatrix{
\rvx_2\ar[d]
&
\rvx_1\ar[l]\ar[d]\ar[dl]
&
\rvx_0\ar[l]\ar[d]\ar[dl]
\\
\rvsig_2
&
\rvsig_1\ar[l]\ar[ul]
&
\rvsig_0\ar[l]\ar[ul]
}
\;.
\label{eq-graph-init}
\eeq
The time reversal of this network must look
like this:

\beq
\entrymodifiers={++[o][F-]}
\xymatrix{
\rvx^*_0\ar[d]
&
\rvx^*_1\ar[l]\ar[d]\ar[dl]
&
\rvx^*_2\ar[l]\ar[d]\ar[dl]
\\
\rvsig^*_0
&
\rvsig^*_1\ar[l]\ar[ul]
&
\rvsig^*_2\ar[l]\ar[ul]
}
\;.
\label{eq-graph-trev}
\eeq
The transition matrices for
each node of the
graph given by
Eq.(\ref{eq-graph-trev})
must be expressible
in some way, yet to be specified,
in terms of the transition matrices
for each node of the
graph given by
Eq.(\ref{eq-graph-init}).
Clearly, if we take
$\rva_j=(\rvs_j,\rvsig_j)$,
then the CB net
given by Eq.(\ref{eq-graph-init})
is a special case of the Markov chain
CB net

\beq
\entrymodifiers={++[o][F-]}
\xymatrix{
\rva_2
&
\rva_1\ar[l]
&
\rva_0\ar[l]
}
\;
\eeq
whose time reversal
network looks like this:

\beq
\entrymodifiers={++[o][F-]}
\xymatrix{
\rva^*_0
&
\rva^*_1\ar[l]
&
\rva^*_2\ar[l]
}
\;.
\eeq
To agree
with our intuition
of how time reversal
should operate, we stipulate that\footnote
{Appendix \ref{app-tr} gives
a specific example of the
time reversal of a Markov chain.}

\beqa
P_{\rva^*_2}(a_2)
&=&
P_{\rva_2}(a_2)
\\
&=&
\sum_{a_1,a_0}P(a_2|a_1)P(a_1|a_0)P(a_0)
\;,
\eeqa

\beqa
P_{\rva^*_1|\rva^*_2}(a_1|a_2)
&=&
 P_{\rva_1|\rva_2}(a_1|a_2)
 \\
&=&
\frac{
\sum_{a_0}
P_{\rva_2|\rva_1}(a_2|a_1)
P_{\rva_1|\rva_0}(a_1|a_0)
P_{\rva_0}(a_0)}
{\sum_{a_1}(num) }
\;,
\eeqa
and

\beqa
P_{\rva^*_0|\rva^*_1}(a_0|a_1)
&=&
 P_{\rva_0|\rva_1}(a_0|a_1)
\\
&=&
\frac{
P_{\rva_1|\rva_0}(a_1|a_0)P_{\rva_0}(a_0)}
{\sum_{a_0}(num) }
\;.
\eeqa

For definiteness, we
will continue to speak of a Markov
chain with only 3 nodes.
Generalization of our statements to the
case of Markov chains with
an arbitrary number of nodes is trivial.

\begin{claim}
\beq
\frac{P_{\rva_0}(a_0)}
{P_{\rva_2}(a_2)}=
\frac{P_{\rva^*_{<3}|\rva^*_2}(a_{<3}|a_2)}
{P_{\rva_{<3}|\rva_0}(a_{<3}|a_0)}
\;.
\eeq
\end{claim}
\proof
\beqa
\frac{P_{\rva^*_{<3}|\rva^*_2}(a_{<3}|a_2)}
{P_{\rva_{<3}|\rva_0}(a_{<3}|a_0)}
&=&
\frac{
P_{\rva^*_0|\rva^*_1}(a_0|a_1)
P_{\rva^*_1|\rva^*_2}(a_1|a_2)
}{
P_{\rva_2|\rva_1}(a_2|a_1)
P_{\rva_1|\rva_0}(a_1|a_0)}
\\
&=&
\frac{
P_{\rva_0|\rva_1}(a_0|a_1)
P_{\rva_1|\rva_2}(a_1|a_2)
}{
P_{\rva_2|\rva_1}(a_2|a_1)
P_{\rva_1|\rva_0}(a_1|a_0)}
\\
&=&
\frac{
P_{\rva_0}(a_0)
}{
P_{\rva_2}(a_2)}
\;.
\eeqa
\qed

Now note that if
we define
$\Sigma$ as
$H(\rvTh_\tau|\rvX_\tau)\trange{0}{\tau}$,
then

\beqa
\Sigma &=&
H(\rvTh_\tau|\rvX_\tau)\trange{0}{\tau}
\\
&=&
\av{
\ln
\frac{
P_{\rvTh_0|\rvX_0}
(\Theta_0|X_0)
}{
P_{\rvTh_\tau|\rvX_\tau}
(\Theta_\tau|X_\tau)
}}_{\Theta_{<3},X_{<3}}
\\
&=&
\av{
\ln
\frac{
P_{\rvTh_0,\rvX_0}
(\Theta_0,X_0)
}{
P_{\rvTh_\tau,\rvX_\tau}
(\Theta_\tau,X_\tau)
}
\frac{
P_{\rvX_\tau}
(X_\tau)
}{
P_{\rvX_0}
(X_0)
}
}_{\Theta_{<3},X_{<3}}
\\
&=&
\av{\hat{\Sigma}}
\;,
\eeqa
where $\hat{\Sigma}$ is defined by

\beq
\hat{\Sigma}=
\ln
\frac{
P_{\rvTh^*_{{<\tau+1}},\rvX^*_{{<\tau+1}}|\rvTh^*_{\tau},\rvX^*_{\tau}}
(\Theta_{{<\tau+1}},X_{{<\tau+1}}|\Theta_{\tau},X_{\tau})
}{
P_{\rvTh_{{<\tau+1}},\rvX_{{<\tau+1}}|\rvTh_{0},\rvX_{0}}
(\Theta_{{<\tau+1}},X_{{<\tau+1}}|\Theta_{0},X_{0})
}
\frac{
P_{\rvX_{{<\tau+1}}|\rvX_{0}}
(X_{{<\tau+1}}|X_{0})
}{
P_{\rvX^*_{{<\tau+1}}|\rvX^*_{\tau}}
(X_{{<\tau+1}}|X_{\tau})
}
\;.
\eeq
In terms
of the operator $\hat{\Sigma}$,
 the CAIN can be stated as

\beq
\av{\hat{\Sigma}}\geq 0
\;.
\label{eq-sig-geq-zero}
\eeq
In analogy to the Jarzynski
equality, Eq.(\ref{eq-sig-geq-zero})
probably generalizes
to

\beq
\av{e^{-\hat{\Sigma}}}= 1
\;.
\label{eq-exp-sig-is-one}
\eeq
Eq.(\ref{eq-exp-sig-is-one})
implies Eq.(\ref{eq-sig-geq-zero})
plus much more.
In fact, if we expand
Eq.(\ref{eq-exp-sig-is-one})
in powers
of $\hat{\Sigma}$,
we get
Eq.(\ref{eq-sig-geq-zero})
from the first order terms
and a fluctuation dissipation
theorem from the second order terms.

This section has considered time
reversal of the CAIN only
for the classical case, but
it can be generalized in a straightforward
way to the quantum case.
To go from the classical to the quantum
case, one replaces CB nets by QB nets,
and probability distributions by
density matrices.
Also classical information functions
$H(\cdot)$ by
quantum information functions $S_\rho(\cdot)$.

\section{Szilard's Engine}

The goal of this section is to apply
Eq.(\ref{eq-lower-bd-minus-df})
to Szilard's heat engine.

Eq.(\ref{eq-lower-bd-minus-df})
gives a lower bound on
the drop
$-dF_\rvX$
in free energy
for a system $\rvX$
that is in contact
with a heat reservoir
$\rvTh$ at temperature
$T$.
The left hand side
of Eq.(\ref{eq-lower-bd-minus-df})
is a sum of two terms,
namely
$-F^\beta(\ham_{\rvX_\tau},
 \rho_{\tau;\rvX_\tau})
\trange{0}{\tau}$
and
$
-TS^\beta(\Delta \ham_{\rvTh_\tau,\rvX_\tau}, \rho_\tau)
\trange{0}{\tau}$,
one
for $\rvX$
and another ``mostly" for
$\rvTh$.
If
we
want to
extract
as much work as possible
 from the system $\rvX$,
we want to make the term for
$\rvTh$,
which is negative, as
close to zero as possible.
So let's assume that the
term for
$\rvTh$ can be made zero.
This means that the thermal variables
must be
``disturbed as little as possible".
According to Eq.(\ref{eq-minus-f-rho-x}), the
term for $\rvX$
is itself
a sum of two terms, namely
$TS(\rho_{\tau;\rvX_\tau})\trange{0}{\tau}$
and
$-
\av{\ham_{\rvX_\tau}}_{\rho_{\tau;\rvX_\tau}}
\trange{0}{\tau}$.
In the case of
the Szilard engine,
the system is an
ideal gas, so
its internal energy
is proportional to the
temperature.
But the temperature
is the same for all $\tau$.
Thus, we shall assume
that the $-
\av{\ham_{\rvX_\tau}}_{\rho_{\tau;\rvX_\tau}}
\trange{0}{\tau}$ term
 is also zero.
This reduces
what
we need
to calculate
for the Szilard
heat engine to
just
the
$TS(\rho_{\tau;\rvX_\tau})\trange{0}{\tau}$
term.
We will calculate this for certain
special forms of the density matrix
$\rho_{\tau;\rvX_\tau}$
that seem good models for
the Szilard engine.

The usual Szilard engine is a
simple version of Maxwell's demon
wherein the
system inside
the box is just one particle.
We will also consider a system
of two particles.
During the cycle of
the engine, a partition is
introduced inside the box,
creating two compartments, and
forcing the particle (or two particles)
to choose sides.
To model this situation, we will use
the following
random variables:

\beq
\underbrace{\rvs,\rvt}_{\rvx},
\underbrace{\rvsig,\rvth}_{\rvd},
\rvb
\;
\eeq

\beq
\underbrace{\rvs,\rvt,\rvsig}_{\rvX},
\underbrace{\rvth,\rvb}_{\rvTh}
\;
\eeq
where

\begin{description}
\item $\rvs=$ system, first particle
\item $\rvt=$ tyro (apprentice), second particle, if
being considered
\item $\rvsig=$ sensor (probe, tape, memory), part of devil
\item $\rvth=$ thermal part of devil, at temperature $T$
\item $\rvb=$ bath at temperature $T$
\item $\rvx=\rvs$ if uni-partite system,
    $\rvx=(\rvs,\rvt)$ if bi-partite system
\item $\rvd=(\rvsig, \rvth)=$ devil.
\item $\rvX=(\rvx,\rvsig)=$
     non-thermal variables (fast changing, not in thermal equilibrium)
\item $\rvTh=(\rvth,\rvb)=$
     thermal variables (slow changing, in thermal equilibrium)
\end{description}

We will consider
four times $\tau=0,1,2,3$, where

\begin{description}
\item $\tau=0$: initial time
\item $\tau=1$: time when
measurement is done,
when system and sensor interact
\item $\tau=2$: time when feedback is done. Information encoded
in the state of the sensor is used to modify the system.
\item $\tau=3$: time when system and sensor are
erased and re-initialized.
\end{description}

We will consider 4 cases: C1, Q1, C2,  Q2,
where
C= classical, Q= quantum, 1= uni-partite system,
2= bi-partite system.

\subsection{C1 Case}\label{sec-c1}

Consider the following CB net

\beq
\entrymodifiers={++[o][F-]}
\xymatrix{
\rvs_3
&
\rvs_2
&
\rvs_1
&
\rvs_0\ar[dl]\ar[l]_>>{\delta}
\\
\rvsig_3
&
\rvsig_2\ar@{}[l]^>>{0}
&
\rvsig_1\ar[lu]\ar[l]_>>{\delta}\ar@{}[r]_>>{0}
&
\rvsig_0
}
\;.
\label{eq-c1-cbnet}
\eeq

In this net:

For the first row of
random variables $\rvs_\tau$:
$P(s_0)$ is arbitrary,
$P(s_1|s_0)=\delta(s_1,s_0)$,
$P(s_2|\sigma_1)$ is arbitrary,
and
$P_{\rvs_3}(s_3)=P_{\rvs_0}(s_3)$.

For the second row of
random variables $\rvsig_\tau$:
$P(\sigma_0)=\delta(\sigma_0,0)$,
$P(\sigma_1|s_0)$ is arbitrary,
$P(\sigma_2|\sigma_1)=\delta(\sigma_2,\sigma_1)$,
and
$P(\sigma_3)=\delta(\sigma_3,0)$.

\begin{figure}[h]
    \begin{center}
    \epsfig{file=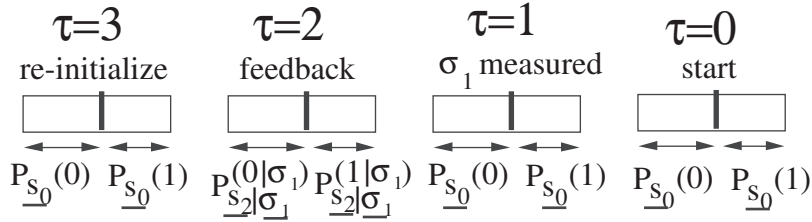, height=1.25in}
    \caption{This
    Szilard engine
    is modeled
    by the CB net given by Eq.(\ref{eq-c1-cbnet}).
    }
    \label{fig-szilard}
    \end{center}
\end{figure}

Fig.\ref{fig-szilard}
shows the position
of the wall of a Szilard
engine with this CB net.

Define
\beq
\Delta H_{vol}=
H(\rvs_0)- H(\rvs_2|\rvsig_1)
\;.
\eeq
The work done by the system
when it changes its volume from
$V_0$ at time $\tau=0$ to
$V_2$ at time $\tau=2$ is
\beq
\ddbar W_{vol}=\int_{V_0}^{V_2}
dV\; P
=\int_{V_0}^{V_2}
dV\; \frac{T}{V}
=
T\ln\frac{V_2}{V_0}
=
T\av{\ln \frac{P(s_2|\sigma_1)}{P(s_0)}}
=
T\Delta H_{vol}
\;.
\eeq
(We assume an ideal
gas so $PV = Nk_B T$ but $N=1$
 and
we are setting $k_B=1$ so $PV = T$)

The following
table is easy to verify
using standard identities
in Shannon Information Theory (specially
the chain rule identities).

\beq
\begin{array}{l|l|l|}
&\hbox{ system } \rvs
& \hbox{ system } \rvs+ \hbox{ sensor } \rvsig
\\
\hline
1\larrow 0
&H(\rvs_\tau)\trange{0}{1}=
&H(\rvs_\tau,\rvsig_\tau)\trange{0}{1}=
\\
&=0
&=-H(\rvsig_1:\rvs_0) + H(\rvsig_1)
\\
\hline
2\larrow 1
&H(\rvs_\tau)\trange{1}{2}=
&H(\rvs_\tau,\rvsig_\tau)\trange{1}{2}=
\\
&=-\Delta H_{vol} + H(\rvs_2:\rvsig_1)
&= -\Delta H_{vol} + H(\rvsig_1:\rvs_0)
\\
\hline
3\larrow 2
&H(\rvs_\tau)\trange{2}{3}=
&H(\rvs_\tau,\rvsig_\tau)\trange{2}{3}=
\\
&=\Delta H_{vol} - H(\rvs_2:\rvsig_1)
&=\Delta H_{vol} - H(\rvsig_1)
\\
\hline
0\larrow 3
&H(\rvs_\tau)\trange{3}{0}=
&H(\rvs_\tau,\rvsig_\tau)\trange{3}{0}=
\\
&=0
&=0
\\
\hline
\end{array}
\;
\label{eq-c1-s-change-table}
\eeq

The entropy change
 over a
full cycle
 is zero, as expected.
For some of the $\tau$, the
entropy change
$H(\rvs_\tau,\sigma_\tau)\trange{0}{\tau}$
contains ``Landauer erasure-work terms" $TH(\sigma_1)$,
``Maxwell volume-work terms" $T\Delta H_{vol}$,
and even
``correlation-energy terms" $TH(\rvsig_1:\rvs_0)$
(these measure a sort of internal energy),
but they all manage to
cancel each other out
over a full cycle.

\subsection{Q1 Case}

In this case, we will abbreviate
$\rho_\tau = \rho_{\tau; \rvs_\tau, \rvsig_\tau}$
and
$S_\tau(\cdot)= S_{\rho_\tau}(\cdot)$.
Also, in this case, $X=(s,\sigma)$.

We begin by
 specifying
 the form of $\rho_\tau$
 that we will assume
 for $\tau=0,1,2,3$.

We will assume that
the sensor
random variable
$\rvsig_\tau$
is a classical random variable
for all $\tau$. Hence,
 $\rvsig_\tau=(\rvsig_\tau)_{cl}$
 for all $\tau$.

\begin{itemize}
\item  At time $\tau=0$,

\beq
\rho_0=
\sum_{r_0}\sum_{\sigma_0}
\sandb{
\sum_{s_0}
\ket{X_0}
A(X_0,r_0)
}
\sandb{\hc}
\;,
\eeq

where

\beq
A(X_0,r_0)=A(s_0,r_0)\delta(\sigma_0,0)
\;,
\eeq
and

\beq
\sum_{s_0,r_0}|A(s_0,r_0)|^2=1
\;.
\eeq

$\rho_0$ can be represented as a QB net
as follows:

\beq
\rho_0=
\begin{array}{l}
{\rm cl}_{\rvsig_0}\\
\tr_{\rvr_0}
\end{array}
\sandb{
\entrymodifiers={++[o][F-]}
\xymatrix@R=6pt@C=6pt{
\rvs_0\ar[r]
&
\rvr_0
\\
\rvsig_0
&
*{}\ar@{}[l]^>>{0}
}
}
\sandb{\hc}
\;.
\eeq

\item At time $\tau=1$,

\beq
\rho_1=
\sum_{r_0}\sum_{\sigma_{1\;:\;0}}
\sandb{
\sum_{s_{1\;:\;0}}
\begin{array}{r}
\ket{X_1}A(X_1|X_0)\\
A(X_0,r_0)
\end{array}
}
\sandb{\hc}
\;,
\eeq
where $A(X_1|X_0)$ is an isometry.

$\rho_1$ can be represented as a QB net
as follows:
\beq
\rho_1=
\begin{array}{l}
{\rm cl}_{\rvsig_1}\\
\tr_{\rvsig_0}\\
\tr_{\rvr_0}
\end{array}
\sandb{
\entrymodifiers={++[o][F-]}
\xymatrix@R=3pt@C=6pt{
\rvs_1
&
*{}
&
\cancel{\rvs_0}\ar[ld]\ar[r]
&
\rvr_0
\\
*{}
&
\scriptstyle \cancel{\rvs_1,\rvsig_1}
\ar[lu]_>>{\delta}\ar[ld]_>>{\delta}
&
*{}
&
*{}
\\
\rvsig_1
&
*{}
&
\rvsig_0\ar[lu]
&
*{}\ar@{}[l]^>>{0}
}
}
\sandb{\hc}
\;.
\eeq

\item  At time $\tau=2$,

\beq
\rho_2=
\sum_{r_0}\sum_{\sigma_{2\;:\;0}}
\sandb{
\sum_{s_{2\;:\;0}}
\begin{array}{r}
\ket{X_2}A(X_2|X_1)\\
A(X_1|X_0)\\
A(X_0,r_0)
\end{array}
}
\sandb{\hc}
\;,
\eeq
where

\beq
A(X_2|X_1)=A(s_2|s_1,\sigma_1)
\delta(\sigma_2,\sigma_1)
\;
\eeq
and,
for all $\sigma_1$,

\beq
\sum_{s_2}
\sandb{A(s_2|s_1,\sigma_1)}
\sandb{\hc\\s_1\rarrow s_1'}=
\delta(s_1,s_1')
\;.
\eeq

$\rho_2$ can be represented as a QB net
as follows:
\beq
\rho_2=
\begin{array}{l}
{\rm cl}_{\rvsig_2}\\
\tr_{\rvsig_{1\;:\;0}}\\
\tr_{\rvr_0}
\end{array}
\sandb{
\entrymodifiers={++[o][F-]}
\xymatrix@R=3pt@C=6pt{
\rvs_2
&
*{}
&
\cancel{\rvs_1}\ar[ld]
&
*{}
&
\cancel{\rvs_0}\ar[ld]\ar[r]
&
\rvr_0
\\
*{}
&
\scriptstyle \cancel{\rvs_2,\rvsig_2}
\ar[lu]_>>{\delta}\ar[ld]_>>{\delta}
&
*{}
&
\scriptstyle \cancel{\rvs_1,\rvsig_1}
\ar[lu]_>>{\delta}\ar[ld]_>>{\delta}
&
*{}
&
*{}
\\
\rvsig_2
&
*{}
&
\rvsig_1\ar[lu]
&
*{}
&
\rvsig_0\ar[lu]
&
*{}\ar@{}[l]^>>{0}
}
}
\sandb{\hc}
\;.
\eeq

\item  At time $\tau=3$,

\beq
\rho_3=
\sum_{R_3, r_3}
\sum_{ r_0}\sum_{\sigma_{3\;:\;0}}
\sandb{
\sum_{s_{3\;:\;0}}
\begin{array}{r}
\ket{X_3}A(X_3,R_3,r_3|X_2)\\
A(X_2|X_1)\\
A(X_1|X_0)\\
A(X_0,r_0)
\end{array}
}
\sandb{\hc}
\;,
\label{eq-rho-3-complicated}
\eeq
where

\beq
A(X_3,R_3,r_3|X_2)=
A_{\rvs_0, \rvr_0}(s_3, r_3)
\delta(\sigma_3,0)A(R_3|s_2,\sigma_2)
\;
\eeq
and $A(R_3|s_2,\sigma_2)$
is an isometry.
Performing the sum over $R_3$,
Eq.(\ref{eq-rho-3-complicated})
 reduces to

\beq
\rho_3=
\sum_{r_3}\sum_{\sigma_3}
\sandb{
\sum_{s_3}
\ket{X_3}
A_{\rvX_0,\rvr_0}(X_3,r_3)
}
\sandb{\hc}
\;.
\eeq

$\rho_3$ can be represented as a QB net
as follows:
\beq
\rho_3=
\begin{array}{l}
{\rm cl}_{\rvsig_3}\\
\tr_{\rvr_3}
\end{array}
\sandb{
\entrymodifiers={++[o][F-]}
\xymatrix@R=6pt@C=6pt{
\rvs_3\ar[r]
&
\rvr_3
\\
\rvsig_3
&
*{}\ar@{}[l]^>>{0}
}
}
\sandb{\hc}
\;.
\eeq
\end{itemize}

For $\tau=1,2$, define\footnote{
In their papers
(Refs.\cite{Sag-U-07}
to \cite{Sag-U-12b}),
Sagawa
and Ueda
introduce a quantity that they
denote by $I_{QC}$
and call the
quantum-classical information.
Their $I_{QC}$
equals our
$\Delta S^{(1)}_{vol}$}

\beq
\Delta S^{(\tau)}_{vol}=
S_0(\rvs_0)- S_\tau(\rvs_\tau|\rvsig_1)
\;
\eeq
and

\beq
\ddbar W^{(\tau)}_{vol}=
T\Delta S^{(\tau)}_{vol}
\;.
\eeq

The following
table is easy to verify
using standard identities
in Shannon Information Theory (specially
the chain rule identities).

\beq
\begin{array}{l|l|l|}
&\hbox{ system } \rvs
& \hbox{ system } \rvs+ \hbox{ sensor } \rvsig
\\
\hline
1\larrow 0
&S_\tau(\rvs_\tau)\trange{0}{1}=
&S_\tau(\rvs_\tau,\rvsig_\tau)\trange{0}{1}=
\\
&=S_1(\rvs_1)-S_0(\rvs_0)
&=-\Delta S^{(1)}_{vol} + H(\rvsig_1)
\\
\hline
2\larrow 1
&S_\tau(\rvs_\tau)\trange{1}{2}=
&S_\tau(\rvs_\tau,\rvsig_\tau)\trange{1}{2}=
\\
&=-\Delta S^{(2)}_{vol} + S_2(\rvs_2:\rvsig_1)
+S_0(\rvs_0)+ S_1(\rvs_1)
&= -\Delta S^{(2)}_{vol} + \Delta S^{(1)}_{vol}
\\
\hline
3\larrow 2
&S_\tau(\rvs_\tau)\trange{2}{3}=
&S_\tau(\rvs_\tau,\rvsig_\tau)\trange{2}{3}=
\\
&=\Delta S^{(2)}_{vol} - S_2(\rvs_2:\rvsig_1)
&= \Delta S^{(2)}_{vol} - H(\rvsig_1)
\\
\hline
0\larrow 3
&S_\tau(\rvs_\tau)\trange{3}{0}=
&S_\tau(\rvs_\tau,\rvsig_\tau)\trange{3}{0}=
\\
&=0
&=0
\\
\hline
\end{array}
\;.
\eeq

\subsection{C2 Case}

In this case,
$x = (s,t)$, and $X=(x,\sigma)$.

Consider the following CB net

\beq
\entrymodifiers={++[o][F-]}
\xymatrix{
\rvs_3,\rvt_3
&
\rvs_2
&
\rvs_1
&
\rvs_0\ar[ddl]\ar[l]_>>{\delta}
&
\rvs_0,\rvt_0\ar[l]_>>{\delta}\ar[dl]_>>{\delta}
\\
*{}
&
\rvt_2
&
\rvt_1
&
\rvt_0\ar[l]_>>{\delta}
&*{}
\\
\rvsig_3
&
\rvsig_2\ar@{}[l]^>>{0}
&
\rvsig_1\ar[lu]\ar[luu]\ar[l]_>>{\delta}
&
\rvsig_0
&*{}\ar@{}[l]^>>{0}
}
\;.
\eeq

In this net:

$P_{\rvs_0,\rvt_0}(s_0, t_0)$ is arbitrary.
$P_{\rvs_3,\rvt_3}(s_3, t_3)
=P_{\rvs_0,\rvt_0}(s_3,t_3)$.

For the first row of
random variables $\rvs_\tau$:
$P(s_0|s'_0, t'_0)=\delta(s_0, s_0')$,
$P(s_1|s_0)=\delta(s_1,s_0)$, and
$P(s_2|\sigma_1)$ is arbitrary.

For the second row of
random variables $\rvt_\tau$:
$P(t_0|s'_0, t'_0)=\delta(t_0, t_0')$,
$P(t_1|t_0)=\delta(t_1,t_0)$, and
$P(t_2|\sigma_1)$ is arbitrary.

For the third row of
random variables $\rvsig_\tau$:
$P(\sigma_0)=\delta(\sigma_0,0)$,
$P(\sigma_1|s_0)$ is arbitrary,
$P(\sigma_2|\sigma_1)=\delta(\sigma_2,\sigma_1)$,
and
$P(\sigma_3)=\delta(\sigma_3,0)$.

Define
\beqa
\Delta H_{vol,\rvs}&=&
H(\rvs_0)- H(\rvs_2|\rvsig_1),
\\
\Delta H_{vol,\rvt}&=&
H(\rvt_0)- H(\rvt_2|\rvsig_1),
\\
\Delta H_{vol,\rvx}&=&
\Delta H_{vol,\rvs} +
\Delta H_{vol,\rvt}
\;.
\eeqa
Let

\beq
\ddbar W_{vol,\rv{\mu}}=
T\Delta H_{vol,\rv{\mu}}
\;
\eeq
for $\rv{\mu}=\rvs,\rvt,\rvx$.

The following
table is easy to verify
using standard identities
in Shannon Information Theory (specially
the chain rule identities).

\beq
\begin{array}{l|l|l|}
&\hbox{ bi-system } \rvx=(\rvs,\rvt)
& \hbox{ bi-system } \rvx=(\rvs,\rvt) + \hbox{ sensor } \rvsig
\\
\hline
1\larrow 0
&H(\rvx_\tau)\trange{0}{1}=
&H(\rvx_\tau,\rvsig_\tau)\trange{0}{1}=
\\
&=0
&=-H(\rvsig_1:\rvs_0) + H(\rvsig_1)
\\
\hline
2\larrow 1
&H(\rvx_\tau)\trange{1}{2}=
&H(\rvx_\tau,\rvsig_\tau)\trange{1}{2}=
\\
&=
\left\{
\begin{array}{l}
-\Delta H_{vol,\rvx} + H(\rvx_2:\rvsig_1)\\
+H(\rvs_0:\rvt_0)
\end{array}
\right.
&=
\left\{
\begin{array}{l}
-\Delta H_{vol,\rvx}+H(\rvsig_1:\rvs_0) \\
+H(\rvs_0:\rvt_0)
\end{array}
\right.
\\
\hline
3\larrow 2
&H(\rvx_\tau)\trange{2}{3}=
&H(\rvx_\tau,\rvsig_\tau)\trange{2}{3}=
\\
&=
\left\{
\begin{array}{l}
\Delta H_{vol,\rvx} - H(\rvx_2:\rvsig_1)\\
-H(\rvs_0:\rvt_0)
\end{array}
\right.
&=
\left\{
\begin{array}{l}
\Delta H_{vol,\rvx} - H(\rvsig_1)\\
-H(\rvs_0:\rvt_0)
\end{array}
\right.
\\
\hline
0\larrow 3
&H(\rvx_\tau)\trange{3}{0}=
&H(\rvx_\tau,\rvsig_\tau)\trange{3}{0}=
\\
&=0
&=0
\\
\hline
\end{array}
\eeq

Note that some of the entropy
changes contain a new kind of term
$TH(\rvs_0:\rvt_0)$, a
``correlation-energy term"
that measures a type of
internal energy of
the bi-partite system.

\subsection{Q2 Case}

In this case, we will abbreviate
$\rho_\tau = \rho_{\tau; \rvs_\tau,
\rvt_\tau,\rvsig_\tau}$
and
$S_\tau(\cdot)=S_{\rho_\tau}(\cdot)$
Also,
in this case, $x = (s,t)$, $X=(x,\sigma)$.

We begin by
 specifying
 the form of $\rho_\tau$
 that we will assume
 for $\tau=0,1,2,3$.
 The form of
 $\rho_\tau$
 is the same as
 that given for the
 Q1 case, except that
 instead of $X=(s,\sigma)$
 we have $X=(s,t,\sigma)$.

For $\tau=1,2$, define
\beqa
\Delta S^{(\tau)}_{vol,\rvs}&=&
S_\tau(\rvs_0)- S_\tau(\rvs_\tau|\rvsig_1),
\\
\Delta S^{(\tau)}_{vol,\rvt}&=&
S_\tau(\rvt_0)- S_\tau(\rvt_\tau|\rvsig_1),
\\
\Delta S^{(\tau)}_{vol,\rvx}&=&
\Delta S^{(\tau)}_{vol,\rvs} +
\Delta S^{(\tau)}_{vol,\rvt}
\;.
\eeqa
Let

\beq
\ddbar W^{(\tau)}_{vol,\rv{\mu}}=
T\Delta S^{(\tau)}_{vol,\rv{\mu}}
\;
\eeq
for $\rv{\mu}=\rvs,\rvt,\rvx$.

The following
table is easy to verify
using standard identities
in Shannon Information Theory (specially
the chain rule identities).

\beq
\begin{array}{l|l|l|}
&\hbox{ bi-system } \rvx=(\rvs,\rvt)
& \hbox{ bi-system } \rvx=(\rvs,\rvt) + \hbox{ sensor } \rvsig
\\
\hline
1\larrow 0
&S_\tau(\rvx_\tau)\trange{0}{1}=
&S_\tau(\rvx_\tau,\rvsig_\tau)\trange{0}{1}=
\\
&=S_1(\rvx_1)-S_0(\rvx_0)
&=-\Delta S^{(1)}_{vol,\rvx} + H(\rvsig_1)
\\
\hline
2\larrow 1
&S_\tau(\rvx_\tau)\trange{1}{2}=
&S_\tau(\rvx_\tau,\rvsig_\tau)\trange{1}{2}=
\\
&=
\left\{
\begin{array}{l}
-\Delta S^{(2)}_{vol,\rvx} + S_2(\rvx_2:\rvsig_1)\\
+S_0(\rvs_0:\rvt_0) + S_0(\rvx_0)-S_1(\rvx_1)
\end{array}
\right.
&=
\left\{
\begin{array}{l}
-\Delta S^{(2)}_{vol,\rvx}+
\Delta S^{(1)}_{vol,\rvx}\\
+S_0(\rvs_0:\rvt_0)
\end{array}
\right.
\\
\hline
3\larrow 2
&S_\tau(\rvx_\tau)\trange{2}{3}=
&S_\tau(\rvx_\tau,\rvsig_\tau)\trange{2}{3}=
\\
&=
\left\{
\begin{array}{l}
\Delta S^{(2)}_{vol,\rvx} - S_2(\rvx_2:\rvsig_1)\\
-S_0(\rvs_0:\rvt_0)
\end{array}
\right.
&=
\left\{
\begin{array}{l}
\Delta S^{(2)}_{vol,\rvx} - H(\rvsig_1)\\
-S_0(\rvs_0:\rvt_0)
\end{array}
\right.
\\
\hline
0\larrow 3
&S_\tau(\rvx_\tau)\trange{3}{0}=
&S_\tau(\rvx_\tau,\rvsig_\tau)\trange{3}{0}=
\\
&=0
&=0
\\
\hline
\end{array}
\eeq
Note that just
as in the C2 case,
here too
some entropy changes
contain ``correlation-energy
terms" $TS_0(\rvs_0:\rvt_0)$
that measure a type of internal energy
of the bi-partite system.

\appendix

\section{Appendix: Very Brief Review
of Pertinent\\
Classical Thermodynamics}\label{app-thermo}

People
with diverse backgrounds
might find the results of this paper
useful. Some of
them
might be rusty or uncomfortable
in their knowledge of
classical thermodynamics. To help those people
out,
here is a brief review of
some facts about classical thermodynamics
that are pertinent to this paper.

As usual, $Q=$ heat,
$E=U=$ internal energy,
$W=$ work,
$P=$ pressure, $V=$ volume,
$S=$ entropy,
$T=$ temperature,
$F=$ free energy.

Let $X$ be  any physical quantity
pertaining to a system.
If $X$ is
an actual function
of the thermodynamical state
of the system (i.e., a ``state function"),
we will use $dX$ to denote a differential,
infinitesimal contribution
to $X$.
If $X$ not a state function, we will use
$\dbar X$ to denote a
non-differential, infinitesimal
contribution to $X$.

We will also use
finite analogues
of $dX$ and $\dbar X$.
If $X$ is a state function,
let $\Delta X$ denote a finite difference,
a finite change
in $X$. If $X$ is not a state function, let
$\ddbar X$ denote a finite contribution
to $X$.

We will also use a subscript of $\cyc$
(for instance, as in $\Delta_\cyc X$)
to indicate that
a change or contribution
occurs over a full cycle
of a cyclic process.

The First Law of
thermodynamics
for a system $s$ is

\beq
\dbar Q_s = dE_s + \dbar W_s
\;.
\eeq
I like to represent it by
a 3-port ``circuit diagram"

\beq
\xymatrix@C=40pt{
\bullet\ar[r]^{\displaystyle\dbar Q_s}_<<{a}
&
\bullet\ar[d]^{\displaystyle dE_s}_>>{b}
\ar[r]^{\displaystyle\dbar W_s}_>>{c}
&
\bullet
\\
*{}&\bullet&*{}
}
\;.
\label{eq-bob-thermo-ckt}
\eeq
When considering more than one
system, one can draw a
3-port circuit
 like Eq.(\ref{eq-bob-thermo-ckt})
for each system.
Given several systems, any pair
of them, say $s_1$ and $s_2$,
might be in thermal contact,
or in mechanical contact.
Thermal contact
(a wall that allows heat to flow across
it from $s_1$ to $s_2$ or vice versa)
can be indicated by
drawing a line
connecting the two $a$ ports  of the
3-port diagrams of $s_1$ and $s_2$.
Mechanical contact
(a wall between
$s_1$ and $s_2$
that is impermeable but free
to move, thus
making the volume of one system
larger and the other smaller)
can be indicated by
drawing a line
connecting  the two $c$ ports
of the 3-port diagrams of $s_1$ and $s_2$.
In a sequence of steps called a ``process", the
thermal and mechanical
contacts can change as a function of time.

Here are some simple
processes often considered
in thermodynamics.

\begin{itemize}
\item[(a)] System $s$ and bath $b$

First Law:
\beq
\xymatrix@C=40pt{
\bullet\ar[r]^{\displaystyle\dbar Q_s}
\ar@{-}[dd]
&
\bullet\ar[d]^{\displaystyle dE_s}
\ar[r]^{\displaystyle\dbar W_s}
&
\bullet
\\
*{}&\bullet&*{}
\\
\bullet\ar[r]^{\displaystyle\dbar Q_b}
&
\bullet\ar[d]^{\displaystyle dE_b}
\ar[r]^{\displaystyle\dbar W_b}
&
\bullet
\\
*{}&\bullet&*{}
}
\;
\eeq

Second Law:

\beq
dS_s + dS_b \geq 0
\;
\eeq

Extra Constraints:

\begin{subequations}
\beq
\dbar Q_b = T dS_b \hbox{  (definition of heat bath)}
\;
\eeq

\beq
\dbar Q_b + \dbar Q_s =0 \hbox{  (thermal contact)}
\;
\eeq

\beq
\xymatrix@C=50pt{
\bullet\ar[r]^{\displaystyle TdS_s}
&
\bullet\ar[d]^{\displaystyle dE_s}
&
\bullet\ar[l]_{dF_s=-PdV}
\\
*{}&\bullet&*{}
}
\;
\eeq
\end{subequations}
(This is a circuit diagram of Eqs.(\ref{eq-df})
and (\ref{eq-du})
at constant temperature).

\begin{claim}
\beq
dS_s \geq \frac{\dbar Q_s}{T}
\;.
\eeq
(Thus, the
entropy of the system
 increases by as much or more
than the
heat/temperaure absorbed by system)
and

\beq
\dbar W_s \leq -dF_s
\;.
\label{eq-2-law-at-t}
\eeq
(Thus, the
drop in free energy of the system
is an upper bound to the amount of
work the system can do.)

For a cycle, $\Delta_\cyc S_s=0$ so
$\ddbar_\cyc Q_s\leq 0$. (No perpetuum mobile of
the first kind.)
\end{claim}
\proof
\beq
0\leq dS_s + dS_b = dS_s + \frac{\dbar Q_b}{T}
= dS_s - \frac{\dbar Q_s}{T}
\;.
\eeq

Eq.(\ref{eq-2-law-at-t})
follows from the following
facts:

\beq
\left\{
\begin{array}{l}
-dF_s=TdS_s-dE_s\\
\dbar W_s = \dbar Q_s - dE_s\\
TdS_s\geq \dbar Q_s
\end{array}
\right.
\;.
\eeq
\qed

\item[(b)] Hot bath $h$ and Cold Bath $c$

First Law:
\beq
\xymatrix@C=40pt{
\bullet\ar[r]^{\displaystyle\dbar Q_h}
\ar@{-}[dd]
&
\bullet\ar[d]^{\displaystyle dE_h}
\ar[r]^{\displaystyle\dbar W_h}
&
\bullet
\\
*{}&\bullet&*{}
\\
\bullet\ar[r]^{\displaystyle\dbar Q_c}
&
\bullet\ar[d]^{\displaystyle dE_c}
\ar[r]^{\displaystyle\dbar W_c}
&
\bullet
\\
*{}&\bullet&*{}
}
\;
\eeq

Second Law:

\beq
dS_h + dS_c \geq 0
\;
\eeq

Extra Constraints:

\begin{subequations}
\beq
\dbar Q_h = T_h dS_h \hbox{  (definition of heat bath)}
\;
\eeq

\beq
\dbar Q_c = T_c dS_c \hbox{  (definition of heat bath)}
\;
\eeq

\beq
\dbar Q_h + \dbar Q_c =0 \hbox{  (thermal contact)}
\;
\eeq

\beq
T_h > T_c
\;.
\eeq
\end{subequations}

\begin{claim}
$\dbar Q_c = -\dbar Q_h \geq 0$.
(Thus, heat flows from hot bath to cold one).
\end{claim}
\proof
\beq
0\leq \frac{\dbar Q_h}{T_h}
+
\frac{\dbar Q_c}{T_c}
=\dbar Q_c\left(
\frac{-1}{T_h}
+
\frac{1}{T_c}
\right)
=\dbar Q_c\left(
\frac{T_h-T_c}{T_h T_c}
\right)
\;.
\eeq
\qed

\item[(c)] Heat Engine

First Law:
\beq
\xymatrix@C=40pt{
\bullet\ar[r]^{\displaystyle \dbar Q_h}
\ar@{-}[dd]
&
\bullet\ar[d]^{\displaystyle dE_h}
\ar[r]^{\displaystyle \dbar W_h}
&
\bullet
\\
*{}&\bullet&*{}
\\
\bullet\ar[r]^{\displaystyle \dbar Q_s}
\ar@{-}[dd]
&
\bullet\ar[d]^{\displaystyle d E_s}
\ar[r]^{\displaystyle \dbar W_s}
&
\bullet
\\
*{}&\bullet&*{}
\\
\bullet\ar[r]^{\displaystyle\dbar Q_c}
&
\bullet\ar[d]^{\displaystyle dE_c}
\ar[r]^{\displaystyle \dbar W_c}
&
\bullet
\\
*{}&\bullet&*{}
}
\;
\eeq

Second Law:

\beq
dS_h + dS_s +
 dS_c = 0
\;
\eeq
(Equality because
assume quasi-static process)
\end{itemize}

Extra Constraints:
\begin{subequations}
\beq
\dbar Q_h = T_h dS_h \hbox{  (definition of heat bath)}
\;
\eeq

\beq
\dbar Q_c = T_c dS_c \hbox{  (definition of heat bath)}
\;
\eeq

\beq
\dbar Q_h + \dbar Q_s
+ \dbar Q_c =0
\hbox{  (thermal contact)}
\;
\eeq

\beq
T_h>T_c
\;
\eeq

\beq
\Delta_\cyc S_s = \Delta_\cyc E_s=0
\hbox{ (one cycle)}
\;
\eeq
\end{subequations}

\begin{claim}
\beq
\frac{\ddbar_\cyc Q_c}{T_c}=
-\left(\frac{\ddbar_\cyc Q_h}{T_h}\right)
\;
\eeq

\beq
\frac{\ddbar_\cyc W_s}
{\ddbar_\cyc Q_c}=
\frac{T_h-T_c}{T_c}=
\mbox{ efficiency}
\;
\eeq
\end{claim}
\proof
\beq
\frac{\ddbar_\cyc Q_c}{T_c}
=\Delta_\cyc S_c
=-\Delta_\cyc S_h
=-\left(\frac{\ddbar_\cyc Q_h}{T_h}\right)
\;.
\eeq

\beq
\ddbar_\cyc W_s
=
\ddbar_\cyc Q_s -\Delta_\cyc E_s
=\ddbar_\cyc Q_s
=-\ddbar_\cyc Q_h -\ddbar_\cyc Q_c
=\left(
\frac{T_h-T_c}{T_c}
\right)
\ddbar_\cyc Q_c
\;
\eeq
\qed

\begin{figure}[h]
    \begin{center}
    \epsfig{file=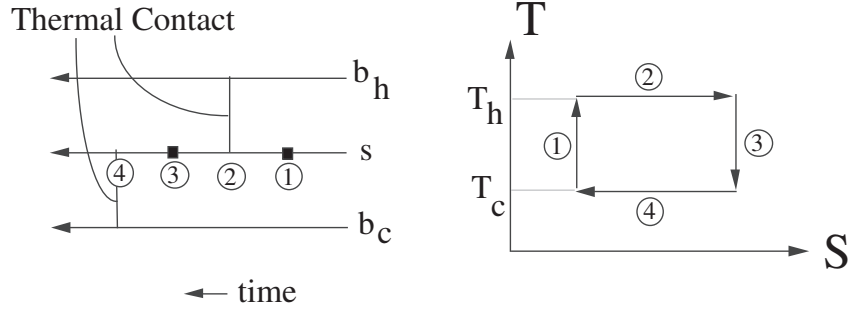, height=1.75in}
    \caption{Cycle of
    a Carnot Engine.
    Steps 1 (baking $s$)
    and 3 (cooling $s$) are isentropic,
    whereas steps 2 and 4 are isothermal.
    }
    \label{fig-carnot}
    \end{center}
\end{figure}

As shown in Fig.\ref{fig-carnot},
the  cycle of a Carnot
engine consists of
a rectangle in the $T,S$ plane.
The system $s$ must first be brought
(via an isentropic, adiabatic step)
to the temperature of
the hot bath $b_h$ (or the cold bath $b_c$),
before it is put in contact with
that bath or else there would
be a temperature difference between
the system and that bath
which would make the process
not quasi-static.

\section{Appendix: Time Reversal of C1 Case}\label{app-tr}
A simple exercise in time reversal
is to find the time reversal of the CB net
given by Eq.(\ref{eq-c1-cbnet}), what
we called the C1 case of Szilard's engine.
Recall $X=(s,\sigma)$.
In our model,

\begin{subequations}
\label{eq-c1-forward}
\beq
P_{\rvX_0}(X_0)=\delta_{\sigma_0}^{0}
P_{\rvs_0}(s_0)
\;,
\eeq

\beq
P_{\rvX_1|\rvX_0}(X_1|X_0)=\delta_{s_0}^{s_1}
P_{\rvsig_1|\rvs_0}(\sigma_1|s_0)
\;,
\eeq
and

\beq
P_{\rvX_2|\rvX_1}(X_2|X_1)=
\delta_{\sigma_2}^{\sigma_1}
P_{\rvs_2|\rvsig_1}(s_2|\sigma_1)
\;.
\eeq
\end{subequations}
Using Eqs.(\ref{eq-c1-forward}),
it is
easy to show
 that

\begin{subequations}
\beq
P_{\rvX^*_2}(X_2)=P_{\rvX_2}(X_2)=
P_{\rvs_2|\rvsig_1}(s_2|\sigma_2)
\sum_{s_0}
P_{\rvsig_1|\rvs_0}(\sigma_2|s_0)P_{\rvs_0}(s_0)
\;,
\eeq

\beq
P_{\rvX^*_1|\rvX^*_2}(X_1|X_2)=
P_{\rvX_1|\rvX_2}(X_1|X_2)=
\delta_{\sigma_2}^{\sigma_1}
\frac{
P_{\rvsig_1|\rvs_0}(\sigma_2|s_1)
P_{\rvs_0}(s_1)
}{
\sum_{s_1}(num)
}
\;,
\eeq
and

\beq
P_{\rvX^*_0|\rvX^*_1}(X_0|X_1)=
P_{\rvX_0|\rvX_1}(X_0|X_1)=
\delta_{s_1}^{s_0}
\delta_{\sigma_0}^{0}
\;.
\eeq
\end{subequations}
Thus the time reversed process has the following
CB net

\beq
\entrymodifiers={++[o][F-]}
\xymatrix{
\rvs^*_0
&
\rvs^*_1\ar[l]_>>{\delta}
&
\rvs^*_2
\\
\rvsig^*_0
&
\rvsig^*_1\ar@{}[l]^>>{0}
&
\rvsig^*_2\ar[l]_>>{\delta}\ar[u]\ar[ul]
}
\;.
\eeq

\section{Appendix: Binary Symmetric Channels}
\label{app-bs-channels}

In this appendix, we will discuss some
of the properties
of binary symmetric channels.
Many results
in classical Shannon
Information Theory
simplify considerably
when they are specialized to
binary symmetric channels.
For instance, the channel capacity
is trivial to calculate
for such channels.\cite{Cover}

Throughout this appendix,
we will assume
 $\alpha,\beta,\gamma,\ell\in[0,1]$.

Define
the complement of $\alpha$ by

\beq
\ol{\alpha}=1-\alpha
\;,
\eeq
and
the symmetric product
of $\alpha$ and $\beta$ by

\beq
\alpha*\beta = \alpha \beta + \ol{\alpha}\;\ol{\beta}
\;.
\eeq

\begin{figure}[h]
    \begin{center}
    \epsfig{file=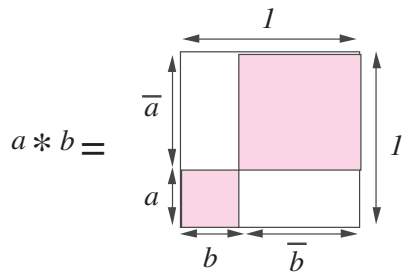, height=1.5in}
    \caption{The symmetric product
    $a*b$ equals the shaded area
    within the unit square.
    }
    \label{fig-sym-prod}
    \end{center}
\end{figure}

As shown in Fig.\ref{fig-sym-prod},
the symmetric product has a simple
geometrical interpretation
in terms of areas
contained in the unit square.

One can easily check that
the symmetric product is commutative
and associative:

\beqa
\alpha*\beta &=& \beta*\alpha\\
\alpha*(\beta*\gamma)&=& (\alpha*\beta)*\gamma
\;.
\eeqa
Other useful properties of
the symmetric product are

\beq
\ol{\alpha}*\beta = 1-\alpha*\beta = \ol{\alpha*\beta}
\;,
\eeq
and

\beqa
\alpha*0&=&\ol{\alpha}\\
\alpha*1 &=& \alpha\\
\alpha*\frac{1}{2} &=& \frac{1}{2}
\;.
\eeqa

Define a symmetric matrix by
\beq
M(\alpha)=
\left[
\begin{array}{cc}
\alpha &\ol{\alpha}\\
 \ol{\alpha} &\alpha
\end{array}
\right]
\;,
\eeq
and a symmetric vector by

\beq
\vec{v}(\ell)=
\left[
\begin{array}{c}
\ell
\\
\ol{\ell}
\end{array}
\right]
\;.
\eeq

One can easily check that
\beq
M(\alpha) \vec{v}(\ell) =
\vec{v}(\alpha*\ell)
\;,
\eeq
and

\beq
M(\beta)M(\alpha) = M(\beta*\alpha)
\;.
\eeq

Define the binary entropy function $h(\alpha)$ by
\beq
h(\alpha)=
-\alpha \ln \alpha
-\ol{\alpha} \ln \ol{\alpha}
\;.
\eeq

A binary symmetric channel
is defined
as the classical Bayesian net
$\rvy\larrow \rvx$,
where the transition matrix $P_{\rvy|\rvx}$
is of the form
\beq
P_{\rvy|\rvx} =
\begin{array}{c|cc}
&\stackrel{x}{\scriptstyle 0} & \stackrel{\rarrow}{\scriptstyle 1}\\
\hline
\scriptstyle \downarrow \;0 & P(y|x)\\
\scriptstyle y\;1&\\
\end{array}=
M(\alpha)
\;.
\eeq
This transition matrix is often
represented by the diagram

\beq
\xymatrix{
0&&0\ar[ll]_<<\alpha\ar[dll]^<<{1-\alpha}\\
1&&1\ar[ll]^<<\alpha\ar[ull]_<<{1-\alpha}
}
\;.
\eeq
Note that the binary symmetric channel
with
$P_{\rvy|\rvx}=M(\alpha)$
is doubly stochastic (the rows and
columns of $M(\alpha)$ sum to one).
It also
satisfies

\beqa
H(\rvy|\rvx)&=&
(P_\rvx(0)+P_\rvx(1))\left(
\alpha \ln \frac{1}{\alpha}
+ \ol{\alpha} \ln \frac{1}{\ol{\alpha}}
\right)
\\
&=&
h(\alpha)
\;.
\eeqa

\begin{claim}
\beq
h(\alpha*\ell)\geq h(\ell)
\;.
\eeq
\end{claim}
\proof
As explained in Ref.\cite{Tuc-ineq},
if $T_{\rvy|\rvx}$ is
a doubly stochastic transition matrix, then
the monotonicity
of the relative entropy
implies that
\beq
H(T_{\rvy|\rvx}P_\rvx)\geq H(P_\rvx)
\;.
\eeq
Now set
$T_{\rvy|\rvx}=M(\alpha)$ and
$P_\rvx=\vec{v}(\ell)$
\qed

Consider the model for
the C1 case of the Szilard engine which
was described in Section \ref{sec-c1}.
Let us specialize that model
by further assuming that
$P_{\rvs_0}=\vec{v}(\ell)$,
$P_{\rvsig_1|\rvs_0}=M(\alpha)$ and
$P_{\rvs_2|\rvsig_1}=M(\beta)$.
Then the table given by
Eq.(\ref{eq-c1-s-change-table})
can be expressed
in terms of the probabilities
$\ell$, $\alpha$ and $\beta$
as follows:

\beq
\begin{array}{l|l|l|}
&\hbox{ system } \rvs
& \hbox{ system } \rvs+ \hbox{ sensor } \rvsig
\\
\hline
1\larrow 0
&H(\rvs_\tau)\trange{0}{1}=
&H(\rvs_\tau,\rvsig_\tau)\trange{0}{1}=
\\
&=0
&=h(\alpha)
\\
\hline
2\larrow 1
&H(\rvs_\tau)\trange{1}{2}=
&H(\rvs_\tau,\rvsig_\tau)\trange{1}{2}=
\\
&=h(\beta*\alpha*\ell)- h(\ell)
&= h(\beta)+h(\alpha*\ell)-h(\alpha)-h(\ell)
\\
\hline
3\larrow 2
&H(\rvs_\tau)\trange{2}{3}=
&H(\rvs_\tau,\rvsig_\tau)\trange{2}{3}=
\\
&=-h(\beta*\alpha*\ell)+h(\ell)
&=-h(\beta)-h(\alpha*\ell)+h(\ell)
\\
\hline
0\larrow 3
&H(\rvs_\tau)\trange{3}{0}=
&H(\rvs_\tau,\rvsig_\tau)\trange{3}{0}=
\\
&=0
&=0
\\
\hline
\end{array}
\eeq

\end{document}